\newtheorem{proposition}{Proposition}
\newtheorem{lemma}{Lemma}

\documentclass[conference]{IEEEtran}

\usepackage{booktabs}
\usepackage{bm}
\usepackage{graphicx}
\usepackage[cmex10]{amsmath}
\usepackage{cite}
\usepackage{amssymb}
\usepackage{subfigure}
\usepackage{extarrows}
\usepackage[letterspace = -2]{microtype}
\usepackage{algorithm}
\usepackage{algpseudocode}
\usepackage{algpascal}
\usepackage{float}
\newfloat{algorithm}{t}{lop}

\begin{document}

\title{
Joint Wireless Information and Power Transfer in a Three-Node Autonomous MIMO Relay Network
}
\author{
\IEEEauthorblockN{Yang Huang\IEEEauthorrefmark{1} and Bruno Clerckx\IEEEauthorrefmark{1}\IEEEauthorrefmark{2}}

\IEEEauthorblockA{\IEEEauthorrefmark{1} Department of Electrical and Electronic Engineering, Imperial College London, London SW7 2AZ, United Kingdom}
\IEEEauthorblockA{\IEEEauthorrefmark{2} School of Electrical Engineering, Korea University, Korea}
\IEEEauthorblockA{Email: \{y.huang13, b.clerckx\}@imperial.ac.uk}
}
\maketitle

\begin{abstract}
This paper investigates a three-node amplify-and-forward (AF) multiple-input multiple-output (MIMO) relay network, where an autonomous relay harvests power from the source information flow and is further helped by an energy flow in the form of a wireless power transfer (WPT) at the destination. An energy-flow-assisted two-phase relaying scheme is proposed, where a source and relay joint optimization is formulated to maximize the rate. By diagonalizing the channel, the problem is simplified to a power optimization, where a relay channel pairing problem is solved by an ordering operation. The proposed algorithm, which iteratively optimizes the relay and source power, is shown to converge. Closed-form solutions can be obtained for the separate relay and source optimizations. Besides, a two-phase relaying without energy flow is also studied. Simulation results show that the energy-flow-assisted scheme is beneficial to the rate enhancement, if the transmit power of the energy flow is adequately larger than that of the information flow. Otherwise, the scheme without energy flow would be preferable.
\end{abstract}


\section{Introduction}
As a promising technology for energy-constrained wireless networks, joint wireless information and power transfer (JWIPT) now attracts much attention in the context of relay networks.

Current research on JWIPT in relay networks mainly studies single-antenna systems with energy-constrained relays\cite{NZDK13, DPEP13, JZ14arXiv} or energy harvesting issues in multiple-antenna systems with power-supplied relays\cite{CMZSA13, LZQ14}. Relaying protocols for wireless-powered relays were firstly proposed in \cite{NZDK13}, which focuses on 3-node single-antenna relay systems by applying the unified power splitting (PS) and time switching (TS) frameworks\cite{ZH13}. More general scenarios of multiple source-destination pairs were studied in \cite{DPEP13}. Besides the above works on single-antenna systems, \cite{CMZSA13} studies a 3-node one-way MIMO relay system where a separated energy harvester extracts wireless power from signals transmitted by a source and a relay. In \cite{LZQ14}, a multiple-antenna relay with power supply is considered. Although \cite{CCL14arxiv} studies a wireless-powered multiple-antenna relay, all other nodes are assumed single-antenna, and the processing matrix at the relay is not optimized. Different from the above research, this paper investigates a one-way amplify-and-forward (AF) multiple-input multiple-output (MIMO) relay network with a wireless-powered relay.

\label{SecSystemModel}
\begin{figure}[!t]
\centering
\includegraphics[width = 2.3in]{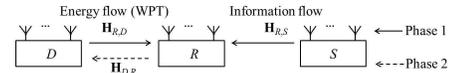}
\caption{JWIET relay network. The destination, relay, and source are designated as $D$, $R$, and $S$, respectively.}
\label{FigRelayingSchemes}
\end{figure}
As shown in Fig. \ref{FigRelayingSchemes}, we consider the scenario where there is no direct link between $S$ and $D$ due to barriers (which causes huge shadow fading), such that the transmission between those two nodes has to rely on a wireless-powered relay $R$. Different from the previous works, we also consider the simultaneous transmission of power (WPT) and information from $D$ and $S$, respectively. In order to efficiently utilize the energy flow (i.e. WPT) but not increase the timeslot consumption, a two-phase energy-flow-assisted relaying scheme is proposed. To make the formulated rate maximization optimization tractable, it is simplified to a power optimization by performing a channel diagonalization based on a harvested-power-maximization power-leakage-minimization (HPM-PLM) strategy. Power allocation at $R$ and $S$ are optimized based on an alternating optimization (AO). Channel pairing issues introduced in the relay power optimization are solved by an ordering operation. Closed-form solutions can be achieved in the separate relay and source power optimizations. While the energy flow provides the relay with an additional source of energy to amplify and forward the information flow, the latter is now subject to the interference from the energy flow. Hence part of the energy harvested at the relay is consumed to amplify and forward the interference, which reduces the power usage effectiveness. An alternative strategy would be to simply rely on a two-phase relaying without (the support of) energy flow. Simulation results indicate that the energy flow-assisted strategy is beneficial to the rate enhancement if the transmit power of the destination is adequately larger than that of the source. Otherwise, the two-phase relaying without energy flow would be preferable.

The remainder of this paper is organized as follows. The system model of the energy-flow-assisted two-phase relaying is formulated in Section \ref{SecSystemModel}. Section \ref{SecChannelDiagnalization} then performs channel diagonalization. Section \ref{SecJointOptimization} discusses the joint relay and source power optimization. Section \ref{SecTwoPhaseRelayingWoEF} elaborates on the two-phase relaying without energy flow scheme. Section \ref{SecPerformanceEvaluation} evaluates the performance of the schemes. Finally, conclusions are drawn in Section \ref{SecConclusion}.

Notations: In this paper, matrices and vectors are in bold capital and bold lower cases, respectively. The notations $\left(\mathbf{A}\right)^T$, $\left(\mathbf{A}\right)^\ast$, $\left(\mathbf{A}\right)^H$, $\text{Tr}\left\{\mathbf{A}\right\}$, $\det\left(\mathbf{A}\right)$, $\lambda_i\left(\mathbf{A}\right)$ and $[\mathbf{A}]_i$ represent the transpose, conjugate, conjugate transpose, trace, determinant, the $i$\,th eigenvalue and the $i$\,th column of a matrix $\mathbf{A}$, respectively. The notation $\mathbf{A} \succeq 0$ means that $\mathbf{A}$ is positive-semidefinite, and $\pi(\mathbf{a})$ and $\|\mathbf{a}\|$ denote the permutation and 2-norm of $\mathbf{a}$, respectively. When $\gtrless$ and $\lessgtr$ are used, top cases or bottom cases in the two notations hold simultaneously.

\section{System Model and Problem Formulation}
\label{SecSystemModel}
In Fig. \ref{FigRelayingSchemes}, each node is equipped with $r$ antennas. The $D$-$R$, $S$-$R$, and $R$-$D$ channels are respectively designated as $\mathbf{H}_{R,D} \in \mathbb{C}^{r \times r}$, $\mathbf{H}_{R,S} \in \mathbb{C}^{r \times r}$, and $\mathbf{H}_{D,R} \in \mathbb{C}^{r \times r}$, which are independent and identically distributed (i.i.d.) Rayleigh flat fading channels, and all the channel matrices are full-rank. Due to channel reciprocity, $\mathbf{H}_{D,R} = \mathbf{H}_{R,D}^T$. Global CSIT is available at each node. The relay exploits PS scheme \cite{ZH13} for simultaneous energy harvesting (EH) and information detecting (ID). At each antenna of the relay, a fraction of the received power, denoted as the PS ratio $\rho_m$ for $m = 1,...,r$, is conveyed to the EH receiver. In this paper, uniform PS is assumed, i.e. $\rho_1, \ldots, \rho_r\ = \rho$. The noise at the ID receiver (at $R$) and $D$ are respectively denoted by $\mathbf{n}_R \sim \mathcal{CN}(0, \sigma_n^2\mathbf{I})$ and $\mathbf{n}_D \sim \mathcal{CN}(0, \sigma_n^2\mathbf{I})$, while the effect of noise at the EH receiver is small and neglected \cite{ZH13, CMZSA13, LZQ14}.

In phase 1, the received signal at the EH receiver is given by $\mathbf{y}_{R,\text{EH}} = \rho^{1/2}\left(\mathbf{H}_{R, D} \mathbf{x}_D + \mathbf{H}_{R, S} \mathbf{x}_S\right)$, where $\mathbf{x}_D$ and $\mathbf{x}_S$ are precoded signals from $D$ and $S$. Assuming an RF-to-DC conversion efficiency of 1, the harvested power equals $\text{Tr}\left\{\rho\mathbf{H}_{R, D} \mathbf{Q}_D \mathbf{H}_{R, D}^H + \rho\mathbf{H}_{R, S} \mathbf{Q}_S \mathbf{H}_{R, S}^H\right\}$, where $\mathbf{Q}_D = \mathcal{E}\{\mathbf{x}_D \mathbf{x}_D^{H}\}$ and $\mathbf{Q}_S = \mathcal{E}\{\mathbf{x}_S \mathbf{x}_S^{H}\}$, respectively. Meanwhile, the baseband signal input to the ID receiver for forwarding is given by $\mathbf{y}_{R,\text{ID}} = (1 - \rho)^{1/2}\left(\mathbf{H}_{R, D} \mathbf{x}_D + \mathbf{H}_{R, S} \mathbf{x}_S\right) + \mathbf{n}_R$. In phase 2, the information received at $D$ is given by
\begin{IEEEeqnarray}{rcl}
\label{EqyARx_P1}
\mathbf{y}_D&=&(1\!-\!\rho)^{1/2}\mathbf{H}_{D, R}\mathbf{F}\left(\mathbf{H}_{R, S} \mathbf{x}_S\!+\!\mathbf{H}_{R, D} \mathbf{x}_D \right)\!+\!\mathbf{n}_R^{\prime} \!+\!\mathbf{n}_D,
\end{IEEEeqnarray}
where $\mathbf{n}_R^{\prime} = \mathbf{H}_{D, R}\mathbf{F}\mathbf{n}_R$ and $\mathbf{F}$ denotes the relay processing matrix. With perfect channel state information, the self-interference in $\mathbf{y}_D$, i.e. the term related to $\mathbf{x}_D$, can be canceled, but some power at the relay is consumed to forward this self-interference. To maximize the achievable rate, an optimization problem can be formulated as
\begin{IEEEeqnarray}{ll}
\text{P1:} & \max_{\mathbf{Q}_D, \mathbf{Q}_S, \mathbf{F}} \frac{1}{2}\log\det \left(\mathbf{I} \! + \! (1 \! - \! \rho)\mathbf{H}_{D, R} \mathbf{F} \mathbf{H}_{R, S} \mathbf{Q}_S \mathbf{H}_{R, S}^H \mathbf{F}^H\cdot\right.\nonumber\\
& \left.\mathbf{H}_{D, R}^H \mathbf{W}^{-1} \right) \IEEEyessubnumber \label{C_NonIF_P1}\\
\text{s.t.} & \text{Tr}\left\{ (1 \! - \! \rho)\left(\mathbf{F}\mathbf{H}_{R, S}\mathbf{Q}_S \mathbf{H}_{R, S}^H \mathbf{F}^H \! + \! \mathbf{F} \mathbf{H}_{R, D} \mathbf{Q}_D \mathbf{H}_{R, D}^H \mathbf{F}^H\right) \right.\nonumber\\
& \left.+ \sigma_n^2\mathbf{F}\mathbf{F}^H \! \right\} \! = \! \rho \text{Tr}\left\{ \! \mathbf{H}_{R, D} \mathbf{Q}_D \mathbf{H}_{R, D}^H \! + \! \mathbf{H}_{R, S}\mathbf{Q}_S \mathbf{H}_{R, S}^H \! \right\},\IEEEyessubnumber \label{FwdPowerConst_1P1} \\
& \text{Tr}\{\mathbf{Q}_D\} \leq P_D \,, \mathbf{Q}_D \succeq 0 \,, \text{Tr}\{\mathbf{Q}_S\} \leq P_S \,, \mathbf{Q}_S \succeq 0\,,\IEEEyessubnumber
\end{IEEEeqnarray}
where the coefficient of $1/2$ in (\ref{C_NonIF_P1}) results from the half-duplex transmission, $\mathbf{W} = \sigma_{n}^2\mathbf{H}_{D, R} \mathbf{F} \mathbf{F}^H \mathbf{H}_{D, R}^H + \sigma_{n}^2\mathbf{I}$, and (\ref{FwdPowerConst_1P1}) implies that all the harvested power at the relay is used for forwarding. For simplicity, $\rho$ is not optimized, but an exhaustive search is conducted to find the best $\rho$. Although the simultaneous transmission in phase 1 is similar to the two-way MIMO relaying\cite{WT12}, optimizing matrices directly as in\cite{WT12} is intractable due to the mutual information criterion (\ref{C_NonIF_P1}) and the constraint (\ref{FwdPowerConst_1P1}). To solve the problem, an iterative algorithm based on channel diagonalization is then proposed.

\section{Channel Diagonalization}
\label{SecChannelDiagnalization}
To simplify the design problem, this section decomposes the forwarding channel $\mathbf{H}_{D, R}$ in phase 2 and jointly decomposes the $S$-$R$ effective channel $\mathbf{\tilde{H}}_{R,S} = \mathbf{H}_{R,S} \mathbf{Q}_S^{1/2}$ and the $D$-$R$ channel $\mathbf{H}_{R, D}$ in phase 1 based on the HPM-PLM strategy, such that  (\ref{C_NonIF_P1}) and (\ref{FwdPowerConst_1P1}) can be diagonalized and problem P1 reduces to a power optimization problem.

\subsection{Structure of Relay Matrix}
As an unique forwarding channel, $\mathbf{H}_{D, R}$ is decomposed as its singular value decomposition (SVD) $\mathbf{U}_{D,R}\mathbf{\Sigma}_{D,R}\mathbf{V}_{D,R}^H$. Then, with the SVD of $\mathbf{\tilde{H}}_{R,S} = \mathbf{\tilde{U}}_{R,S} \mathbf{\tilde{\Sigma}}_{R,S}\mathbf{\tilde{V}}_{R,S}^H$, applying the matrix inversion lemma to (\ref{C_NonIF_P1}) yields
\begin{IEEEeqnarray}{rcl}
\label{C_NonIF_P1_D}
C &{}={}& \frac{1}{2}\log\det \left(\mathbf{I} + \frac{(1 - \rho)}{\sigma_n^2} \mathbf{\tilde{\Sigma}}_{R,S} \left(\mathbf{I} - \left(\mathbf{I} + \right.\right.\right.\nonumber\\
&& \left.\left.\left. \mathbf{\tilde{U}}_{R,S}^H \mathbf{F}^H \mathbf{H}_{D,R}^H \mathbf{H}_{D,R} \mathbf{F} \mathbf{\tilde{U}}_{R,S} \right)^{-1}\right) \mathbf{\tilde{\Sigma}}_{R,S} \right) \,,
\end{IEEEeqnarray}
where the matrix between the two $\mathbf{\tilde{\Sigma}}_{R,S}$ equals a positive semidefinite matrix $\mathbf{\tilde{U}}_{R,S}^H \mathbf{F}^H\mathbf{H}_{D,R}^H(\mathbf{I} + \mathbf{H}_{D,R} \mathbf{F} \mathbf{F}^H \mathbf{H}_{D,R}^H)^{-1} \cdot \mathbf{H}_{D,R} \mathbf{F} \mathbf{\tilde{U}}_{R,S}$. Hence, the matrix in $\det(\cdot)$ is positive-definite. According to Hadamard's inequality \cite{CT91}, (\ref{C_NonIF_P1_D}) is maximized provided $\mathbf{\tilde{U}}_{R,S}^H \mathbf{F}^H \mathbf{V}_{D,R} \mathbf{\Sigma}_{D,R}^2 \mathbf{V}_{D,R}^H \mathbf{F} \mathbf{\tilde{U}}_{R,S}$ is diagonal. Hence, $\mathbf{F}\!=\!\mathbf{V}_{D,R} \mathbf{\Sigma}_F \mathbf{\tilde{U}}_{R,S}^H$ for $\mathbf{\Sigma}_F\!\in\!\mathbb{C}^{r \times r}$, which means that the relay couples a given receive eigenmode of $\mathbf{\tilde{U}}_{R,S}$ with a given transmit eigenmode of $\mathbf{V}_{D,R}$ with an amplification factor given by the corresponding diagonal entry of $\mathbf{\Sigma}_F$. With the decomposed $\mathbf{F}$, (\ref{C_NonIF_P1}) can be diagonalized.

\subsection{Maximize Harvested Power and Minimize Power Leakage}
\label{SecHarvestedPowerMaximizationPowerLeakageMinimization}
The HPM-PLM strategy is proposed to enhance the power usage effectiveness at $R$, since the energy flow harvested at the EH receiver is used to forward not only the information but also the energy flow leaking into the ID receiver, as shown in (\ref{FwdPowerConst_1P1}). Performing eigenvalue decompositions (EVD), $\mathbf{Q}_D = \mathbf{V}_D \mathbf{\Sigma}_D^2 \mathbf{V}_D^H$ and $\mathbf{H}_{R,D} \mathbf{Q}_D  \mathbf{H}_{R,D}^H\! = \!\mathbf{\tilde{U}}_{R,D} \mathbf{\tilde{\Sigma}}_{R,D}^2 \mathbf{\tilde{U}}_{R,D}^H$. Given the SVD of $\mathbf{H}_{R,D}\!=\!\mathbf{V}_{D,R}^{\ast} \mathbf{\Sigma}_{D,R} \mathbf{U}_{D,R}^T$, rearranging (\ref{FwdPowerConst_1P1}) yields
\begin{IEEEeqnarray}{cl}
\label{FwdPowerConst_2}
&\text{Tr}\left\{(1 - \rho) \mathbf{\Sigma}_F^H \mathbf{\Sigma}_F \mathbf{\tilde{\Sigma}}_{R,S}^2 + \sigma_n^2\mathbf{\Sigma}_F^H \mathbf{\Sigma}_F - \rho \mathbf{\tilde{\Sigma}}_{R,S}^2\right\} \label{FwdPowerConst_2_1} \\
{}={}&\text{Tr}\left\{ \left(\rho \mathbf{I} - (1 - \rho)\mathbf{\Sigma}_F^H \mathbf{\Sigma}_F\right) \left(\mathbf{\tilde{U}}_{R,S}^H \mathbf{V}_{D,R}^{\ast} \mathbf{\Sigma}_{D,R} \mathbf{U}_{D,R}^T \mathbf{V}_D \cdot \right.\right.\nonumber \\
&\left.\left.\mathbf{\Sigma}_D^2 \mathbf{V}_D^H \mathbf{U}_{D,R}^{\ast} \mathbf{\Sigma}_{D,R} \mathbf{V}_{D,R}^T \mathbf{\tilde{U}}_{R,S}\right)  \right\} \label{FwdPowerConst_2_2}\\
{}={}& \! \text{Tr} \! \left\{ \! \left(\rho \mathbf{I} \! - \! (1 \! - \! \rho)\mathbf{\Sigma}_F^H \mathbf{\Sigma}_F \! \right) \! \left( \! \mathbf{\tilde{U}}_{R,S}^H \! \mathbf{\tilde{U}}_{R,D} \! \mathbf{\tilde{\Sigma}}_{R,D}^2 \! \mathbf{\tilde{U}}_{R,D}^H \! \mathbf{\tilde{U}}_{R,S} \! \right) \! \right\}\!.\label{FwdPowerConst_2_3}
\end{IEEEeqnarray}

Eq. (\ref{FwdPowerConst_2_3}) highlights the difference between the power of the energy flow harvested at the EH receiver and the energy flow leaking into the ID receiver. Thus, a strategy can be proposed to maximize the harvested energy and minimize the power leakage. To maximize the transfer of energy from $D$ to $R$, rank-one transmission should be exploited at $D$, i.e. $\mathbf{Q}_D \! =\! P_D [\mathbf{U}_{D,R}^{\ast}]_{\text{max}} [\mathbf{U}_{D,R}^{\ast}]_{\text{max}}^H$, where $[\mathbf{U}_{D,R}^{\ast}]_{\text{max}}$ is the right singular vector (RSV) corresponding to the maximum singular value $\lambda_{D,R,\text{max}}^{1/2}$ of $\mathbf{H}_{R,D}$, such that $\rho \text{Tr}\{ \! \mathbf{H}_{R, D} \mathbf{Q}_D \mathbf{H}_{R, D}^H \!\} \! = \!\rho P_D\lambda_{D,R,\text{max}}$ (see Proposition 1 in \cite{PC13} for proof). To minimize the power leakage, the power leaking into the ID receiver should be paired with the minimum amplification coefficient, such that the power of the retransmitted leakage equals $(1 - \rho)\lambda_{f,\text{min}} P_D \lambda_{D,R,\text{max}}$, where $\lambda_{f,\text{min}}$ denotes the minimum diagonal entry of $\mathbf{\Sigma}_F^H \mathbf{\Sigma}_F$. With the above strategy, (\ref{FwdPowerConst_2_3}) should equal $(\rho - (1 - \rho)\lambda_{f,\text{min}}) P_D \lambda_{D,R,\text{max}}$, which is shown to be an upper bound of (\ref{FwdPowerConst_2_3}) by applying Lemma II.1 in \cite{Lasserre95}. To make (\ref{FwdPowerConst_2_3}) equal to the upper bound, in (\ref{FwdPowerConst_2_2}), $\mathbf{\tilde{U}}_{R,S}^H \mathbf{V}_{D,R}^{\ast} = \mathbf{P_\pi}$. This specific $\mathbf{P_\pi}$ permutates the unique non-zero diagonal entry $P_D \lambda_{D,R,\text{max}}$ in the diagonal matrix $\mathbf{\Sigma}_{D,R} \mathbf{U}_{D,R}^T \mathbf{V}_D \mathbf{\Sigma}_D^2 \mathbf{V}_D^H \mathbf{U}_{D,R}^{\ast} \mathbf{\Sigma}_{D,R}$ to the same position as $\rho - (1 - \rho)\lambda_{f,\text{min}}$ in $\rho \mathbf{I} \! - \! (1 \! - \! \rho)\mathbf{\Sigma}_F^H \mathbf{\Sigma}_F$. By this means, (\ref{FwdPowerConst_2_3}) achieves the upper bound, and (\ref{FwdPowerConst_1P1}) is diagonalized.

In summary, (\ref{C_NonIF_P1}) is diagonalized with the decomposed $\mathbf{H}_{D,R}$ and the structure of $\mathbf{F}$, and (\ref{FwdPowerConst_1P1}) is diagonalized with the HPM-PLM strategy, i.e. the rank-one $\mathbf{Q}_D$ and $\mathbf{\tilde{U}}_{R,S}\!=\!\mathbf{V}_{D,R}^{\ast}\mathbf{P}_{\mathbf{\pi}}^T$. \mbox{Since} $\mathbf{Q}_S = \mathbf{H}_e \mathbf{\tilde{\Sigma}}_{R,S}^2 \mathbf{H}_e ^{H}$ where $\mathbf{H}_e = (\mathbf{\tilde{U}}_{R,S}^H \cdot \mathbf{H}_{R, S})^{-1}$, $\text{Tr}\{\mathbf{Q}_S\} = \sum_{m = 1}^r \|\mathbf{h}_{e,m}\|^2 \tilde{\lambda}_{R,S,m} \leq P_S$, where $\mathbf{h}_{e,m} = [\mathbf{H}_e]_m$ and $\tilde{\lambda}_{R,S,m}$ denotes the $m$\,th diagonal entry of $\mathbf{\tilde{\Sigma}}_{R,S}^2$. Hence, problem P1 reduces to a power optimization.
Since $\mathbf{F}  = \mathbf{V}_{D,R} \mathbf{\Sigma}_F \mathbf{P}_{\mathbf{\pi}}\mathbf{V}_{D,R}^T$, the RSV of $\mathbf{F}$ (i.e. $\mathbf{P}_{\mathbf{\pi}} \mathbf{V}_{D,R}^T$) matches the left singular vectors (LSV) of $\mathbf{\tilde{H}}_{R,S}$ and a permutation of the LSV of $\mathbf{H}_{D, R}$. According to (\ref{EqyARx_P1}), $\mathcal{E}\{\mathbf{y}_D \mathbf{y}_D^H\} = \mathbf{U}_{D,R}[\mathbf{\Sigma}_{D,R}^2\mathbf{\Sigma}_F\mathbf{\Sigma}_F^H(1 \! -\! \rho) (\mathbf{\tilde{\Sigma}}_{R,S}^2 \! + \! \mathbf{P}_{\mathbf{\pi}} \mathbf{\Sigma}_{D,R}^2 \mathbf{U}_{D,R}^T \mathbf{Q}_D\mathbf{U}_{D,R}^\ast \mathbf{P}_{\mathbf{\pi}}^T)  +  (\mathbf{\Sigma}_{D,R}^2\mathbf{\Sigma}_F\mathbf{\Sigma}_F^H \! + \! 1)\sigma_n^2 ]\mathbf{U}_{D,R}^H$, where $\mathbf{U}_{D,R}^T\mathbf{Q}_D\mathbf{U}_{D,R}^\ast$ is diagonal. In the above $\mathcal{E}\{\mathbf{y}_D \mathbf{y}_D^H\}$, because of the same (but permutated) LSV of $\mathbf{\tilde{H}}_{R,S}$ and $\mathbf{H}_{R,D}$, the channel power gains of the effective $S$-$R$ and $D$-$R$ channels in phase 1 are overlapped at the ID receiver, i.e. $(1\!-\!\rho) (\mathbf{\tilde{\Sigma}}_{R,S}^2 \! + \! \mathbf{P}_{\mathbf{\pi}} \mathbf{\Sigma}_{D,R}^2  \mathbf{U}_{D,R}^T \mathbf{Q}_D \mathbf{U}_{D,R}^\ast  \mathbf{P}_{\mathbf{\pi}}^T)$, where the diagonal entries of $\mathbf{\tilde{\Sigma}}_{R,S}^2$ and $(1 - \rho)\mathbf{P}_{\mathbf{\pi}} \mathbf{\Sigma}_{D,R}^2 \mathbf{U}_{D,R}^T \mathbf{Q}_D \mathbf{U}_{D,R}^\ast \mathbf{P}_{\mathbf{\pi}}^T$ are respectively denoted as $\mathbf{\tilde{\lambda}}_{R,S} \triangleq [\tilde{\lambda}_{R,S,1}, \ldots, \tilde{\lambda}_{R,S,r}]^T$ and $\mathbf{\beta} \triangleq [\beta_1, \ldots, \beta_r]^T$ with the unique non-zero $\beta_m\!=\!(1\!-\!\rho)P_D \lambda_{D,R,\text{max}} \triangleq c$. Although the retransmitted energy flow can be canceled at $D$ (i.e. $\mathbf{P}_{\mathbf{\pi}} \mathbf{\Sigma}_{D,R}^2 \mathbf{U}_{D,R}^T \mathbf{Q}_D\mathbf{U}_{D,R}^\ast \mathbf{P}_{\mathbf{\pi}}^T$ in $\mathcal{E}\{\mathbf{y}_D \mathbf{y}_D^H\}$ is deleted), the overlapped channel power gains in phase 1 still impact the rate, because they are amplified and transmitted. As shown in the diagonalized relay power constraint (\ref{FwdPowerConst_1P1}) (i.e. the following (\ref{EqEqConstP3a})), the diagonal entries of $\mathbf{\Sigma}_F\mathbf{\Sigma}_F^H$, denoted as $\mathbf{\lambda}_f \triangleq [\lambda_{f,1}, \ldots, \lambda_{f,r}]^T$, are non-uniformly weighted by the entries of the overlapped channel power gains i.e. $(1 - \rho)\tilde{\lambda}_{R,S,m} + \beta_m$ for $m = 1,\ldots, r$. Thus, the pairings of the diagonal entries of $\mathbf{\Sigma}_{D,R}^2$ (related to the forwarding eigenmodes in phase 2) and the overlapped channel power gains in phase 1 affect the optimization of $\mathbf{\lambda}_f$ and thereby the rate. Additionally, in phase 1, the value of each $(1 - \rho)\tilde{\lambda}_{R,S,m} + \beta_m$ is affected by $\mathbf{P}_{\mathbf{\pi}}$ (which determines the pairing of each $\tilde{\lambda}_{R,S,m}$ and $\beta_m$).

\section{Joint Power Allocation Optimization}
\label{SecJointOptimization}
To make further calculation and analysis tractable, we then focus on the achievable rate at high receive SNR. Substituting the previous channel decompositions into problem P1, the original problem can be reformulated as
\begin{IEEEeqnarray}{ll}
\text{P2:}&\min_{\mathbf{\lambda}_f,\mathbf{\tilde{\lambda}}_{R,S}} -\sum_{m = 1}^{r} \log \left(\frac{(1 - \rho) \tilde{\lambda}_{R,S,m} \lambda_{f,m} \lambda_{D,R,m}}{\sigma_n^2\left(1 + \lambda_{f,m} \lambda_{D,R,m}\right)}\right) \IEEEyessubnumber \label{C_NonIF_Scalar_P2} \\
\text{s.t.} & \lambda_{f,1}, \lambda_{f,2}, \ldots, \lambda_{f,r} > 0\,, \IEEEyessubnumber \label{EqldaFineqConst}\\
& \tilde{\lambda}_{R,S,1}, \tilde{\lambda}_{R,S,2}, \ldots, \tilde{\lambda}_{R,S,r} > 0\,, \IEEEyessubnumber \label{EqldaRSineqConst} \\
& \text{Tr}\{\mathbf{Q}_S\} = \sum_{m = 1}^r \|\mathbf{h}_{e,m}\|^2 \tilde{\lambda}_{R,S,m} \leq P_S,\IEEEyessubnumber \label{EqSrcPwrConst} \\
& \sum_{m = 1}^r \lambda_{f,m} \left( (1 - \rho)\tilde{\lambda}_{R,S,m} + \sigma_n^2 + \beta_m\right)\!=\!\sum_{m = 1}^r \rho \tilde{\lambda}_{R,S,m} +\nonumber\\
& \rho P_D \lambda_{D,R,\text{max}} \,, \IEEEyessubnumber \label{EqEqConstP3a}
\end{IEEEeqnarray}
where $\beta_m$ is constrained by $\beta_m\!=\!c$, if $m\!=\!\text{index}(\lambda_{f,\text{min}})$ (where $\text{index}(\!\lambda_{f,\text{min}}\!)$ returns the index of $\lambda_{f,\text{min}}$); otherwise, $\beta_m =0$. Problem P2 is not convex due to the non-affine (\ref{EqEqConstP3a}). Then, problem P2 is solved using an AO.

\subsection{Relay Optimization with Fixed Source Power Allocation}
\label{SecOptimRelayPwrAllocWithFixedSrcPwrAlloc}
With given $\tilde{\lambda}_{R,S,m}$, the power optimization problem at the relay is formulated as
\begin{IEEEeqnarray}{cl}
\text{P3(a):}\, & \max_{\mathbf{\lambda}_f}  \sum_{m = 1}^{r} \log \left( \frac{(1 - \rho) \tilde{\lambda}_{R,S,m} \lambda_{f,m} \lambda_{D,R,m}}{\sigma_n^2\left(1 + \lambda_{f,m} \lambda_{D,R,m}\right)}\right) \label{EqObjFuncP3a} \\
\text{s.t.} & \text{(\ref{EqldaFineqConst}) and (\ref{EqEqConstP3a})}\,.\nonumber
\end{IEEEeqnarray}
The challenge in solving P3(a) is that $c$ of $\beta_m$ is required to be paired with $\lambda_{f,\text{min}}$, but the position of $\lambda_{f,\text{min}}$ in $\mathbf{\lambda}_f$ is unknown before solving the problem; the rate is affected by the pairings of the elements of $\mathbf{\lambda}_{D,R}\triangleq [\lambda_{D,R,1}, \ldots, \lambda_{D,R,r}]^T$ (i.e. the diagonal entries of $\mathbf{\Sigma}_{D,R}^2$ related to the forwarding eigenmodes in phase 2) and $\mathbf{\tilde{\lambda}}_{R,S}$ (related to $\mathbf{\tilde{H}}_{R,S}$ in phase 1), even if the constraint on $c$ is relaxed and $\beta_m$ is fixed. To avoid the high complexity of searching the best pairings for P3(a), we then reveal that the pairing issues can be solved by ordering operations. Firstly, we relax the constraint on $c$, i.e. $\mathbf{P}_{\mathbf{\pi}}$ becomes an arbitrary permutation matrix $\mathbf{\tilde{P}}_{\mathbf{\pi}}$, and assume columns of $\mathbf{V}_{D,R}$ are arranged in certain orders, such that the elements of $\mathbf{\lambda}_{D,R}$ and $\mathbf{\tilde{\lambda}}_{R,S}$ are paired in certain ways and $c$ is paired with a certain $\tilde{\lambda}_{R,S,m}$. Problem P3(a) then becomes convex regardless of the pairing issues. By analyzing Karush-Kuhn-Tucker (KKT) conditions, a closed-form solution can be obtained by
\begin{IEEEeqnarray}{l}
\label{EqClosedSoluP3a}
\lambda^{\star}_{f,m} = - \frac{1}{2\lambda_{D,R,m}}\nonumber\\
\, + \frac{1}{2} \! \sqrt{\frac{1}{\lambda^2_{D,R,m}} \! + \! \frac{4}{\nu^{\star}\lambda_{D,R,m}\left(\!(1 \! - \! \rho)\tilde{\lambda}_{R,S,m} \! + \! \sigma_n^2 \! + \! \beta_m \right)}},
\end{IEEEeqnarray}
where $\nu^{\star}$ denotes the Lagrange multiplier for the constraint (\ref{EqEqConstP3a}) and is greater than 0. It can be calculated by solving $\sum^r_{m = 1} \lambda^{\star}_{f,m}\left((1\!-\!\rho)\tilde{\lambda}_{R,B,m}\!+\!\sigma_n^2 + \beta_m \right)\!=\!\rho P_D \lambda_{D,R,\text{max}} + \sum^r_{m = 1}\rho\tilde{\lambda}_{R,S,m}$ with bisection. By using (\ref{EqClosedSoluP3a}), two lemmas are revealed. For notational simplicity, it is defined that $z_m \triangleq (1 - \rho)\tilde{\lambda}_{R,S,m} + \sigma_n^2 + \beta_m$, $\mathbf{z} \triangleq [z_1, \ldots, z_r]^T$; $l_m \triangleq (1 - \rho)\tilde{\lambda}_{R,S,m} + \sigma_n^2$, $\mathbf{l} \triangleq [l_1, \ldots, l_r]^T$.

\begin{lemma}
\label{Lemma1}
Suppose that the elements in $\pi_1(\mathbf{z})$ are arranged in the same order as another permutation $\pi_2(\mathbf{z})$ except that $z_i$ and $z_j$ (where $z_i \leq z_j$ for $i < j$) in $\pi_1(\mathbf{z})$ are swapped in $\pi_2(\mathbf{z})$, i.e., $z_i$ and $z_j$ in $\pi_1(\mathbf{z})$ are paired with $\lambda_{D,R,p}$ and $\lambda_{D,R,q}$ (where $\lambda_{D,R,p} \leq \lambda_{D,R,q}$ for $p < q$), respectively, while $z_j$ and $z_i$ in $\pi_2(\mathbf{z})$ are paired with $\lambda_{D,R,p}$ and $\lambda_{D,R,q}$, respectively. Then, the value of the objective function (\ref{EqObjFuncP3a}) with $\pi_1(\mathbf{z})$ is no less than that with $\pi_2(\mathbf{z})$.
\end{lemma}
\begin{IEEEproof}
To prove this lemma, $\pi_1(\mathbf{z})$ and $\pi_2(\mathbf{z})$ are respectively substituted into (\ref{EqClosedSoluP3a}) with $\mathbf{\lambda}_{D,R}$ to calculate and compare the values of (\ref{EqObjFuncP3a}). The lemma is finally justified by scaling Inequalities. See our extended version \cite{HC14JWIETrelayJournalV} for details.
\end{IEEEproof}

Lemma \ref{Lemma1} addresses the pairings of the transmit eigenmodes of $\mathbf{F}$ (i.e. $\mathbf{V}_{D,R}$) and the overlapped channel power gains. With fixed pairings of $\tilde{\lambda}_{R,S,m}$ and $\beta_m$ for $m = 1\ldots r$, the values of the entries of the overlapped channel power gains, i.e. $(1-\rho)\tilde{\lambda}_{R,S,m} + \beta_m$ for $m = 1\ldots r$, are fixed. Lemma \ref{Lemma1} reveals that, for two pairs of the transmit eigenmodes of $\mathbf{V}_{D,R}$ and the overlapped channel power gains (while other pairings are fixed), the strongest eigenmode of $\mathbf{V}_{D,R}$ and the strongest overlapped channel power gain should be paired together. The following Lemma \ref{Lemma2} addresses the pairings of the channel power gains of the effective $S$-$R$ channel and the non-zero channel power gain of the effective $D$-$R$ channels, i.e. the pairings of $\tilde{\lambda}_{R,S,m}$ and $c$ for $m = 1 \ldots r$. It is shown that, for two channel power gains in $\tilde{\lambda}_{R,S,m}$, the strongest $\tilde{\lambda}_{R,S,m}$ should be paired with $c$.

\begin{lemma}
\label{Lemma2}
Assume two permutations $\pi_1(\mathbf{l})$ and $\pi_2(\mathbf{l})$. In $\pi_1(\mathbf{l})$, positions of $l_i$ and $l_j$ follows that $\min\{l_i + c, l_j\}$ and $\max\{l_i + c, l_j\}$ are respectively paired with $\lambda_{D,R,i}$ and $\lambda_{D,R,j}$ (where $i < j$, $l_i \leq l_j$ and $\lambda_{D,R,i} \leq \lambda_{D,R,j}$). In $\pi_2(\mathbf{l})$, positions of $l_i$ and $l_j$ follows that $\min\{l_i, l_j + c\}$ and $\max\{l_i, l_j + c\}$ are respectively paired with $\lambda_{D,R,i}$ and $\lambda_{D,R,j}$. Other pairings between $\lambda_{D,R,m}$ and $l_m$ (for $m \neq i,j$) in $\pi_1(\mathbf{l})$ are the same as $\pi_2(\mathbf{l})$. Then, the objective function (\ref{EqObjFuncP3a}) with $c$ paired with $l_j$ yields a higher value than that with $c$ paired with $l_i$.
\end{lemma}
\begin{IEEEproof}
Based on the conclusion of Lemma \ref{Lemma1}, Lemma \ref{Lemma2} is proved. The idea to prove this lemma is similar to Lemma \ref{Lemma1}. See \cite{HC14JWIETrelayJournalV} for details.
\end{IEEEproof}
\begin{proposition}
\label{Proposition1}
When the elements in $\mathbf{\tilde{\lambda}}_{R,S}$ and $\mathbf{\lambda}_{D,R}$ are arranged in increasing orders and the non-zero $\beta_m$ is paired with the maximum $\tilde{\lambda}_{R,S,m}$, the value of the objective function (\ref{EqObjFuncP3a}) in Problem P3(a) is maximized and the optimal $\lambda^{\star}_{f,m}$ are arranged in a decreasing order.
\end{proposition}
\begin{IEEEproof}
Applying Lemma \ref{Lemma1} and Lemma \ref{Lemma2}, the orderings of $\mathbf{\tilde{\lambda}}_{R,S}$ and $\mathbf{\lambda}_{D,R}$ and the pairing of $\beta_m$ and $\tilde{\lambda}_{R,S,m}$ in the proposition are proved through induction. Based on the above orderings and pairing, the ordering of $\lambda^{\star}_{f,m}$ can be easily proved. See the Proposition \ref{Proposition1} in \cite{HC14JWIETrelayJournalV} for details.
\end{IEEEproof}

Proposition \ref{Proposition1} illustrates that the constraint on $\beta_m$ (i.e. $c$ is paired with $\lambda_{f,\text{min}}$) can be safely relaxed. Following the ordering operation in Proposition \ref{Proposition1}, entries $(\rho - (1 - \rho)\lambda_{f,\text{min}})$ and $P_D \lambda_{D,R,\text{max}}$ are at lower-right corners of matrices $\rho \mathbf{I} - (1 - \rho) \mathbf{\Sigma}_F^H \mathbf{\Sigma}_F$ and $\mathbf{\Sigma}_{D,R} \mathbf{U}_{D,R}^T \mathbf{V}_D \mathbf{\Sigma}_D^2  \mathbf{V}_D^H  \mathbf{U}_{D,R}^{\ast} \mathbf{\Sigma}_{D,R}$ in (\ref{FwdPowerConst_2}), respectively. Hence, the permutation matrix $\mathbf{\tilde{P}}_{\mathbf{\pi}} = \mathbf{I} = \mathbf{P}_{\mathbf{\pi}}$.

\subsection{Source Optimization with Fixed Relay Power Allocation}
According to Proposition \ref{Proposition1}, $\lambda_{f,m}$ are arranged in a decreasing order, and $\text{index}(\lambda_{f,\text{min}}) =r$. Thus, the source power optimization problem is formulated as
\begin{IEEEeqnarray}{cl}
\text{P3(b):}\,& \min_{\mathbf{\tilde{\lambda}}_{R,S}}- \! \sum_{m = 1}^{r} \log \left( \frac{(1 - \rho) \tilde{\lambda}_{R,S,m} \lambda_{f,m} \lambda_{D,R,m}}{\sigma_n^2\left(1 + \lambda_{f,m} \lambda_{D,R,m}\right)}\right) \IEEEyessubnumber \label{EqObjFuncP3b} \\
\text{s.t.} & 0 < \tilde{\lambda}_{R,S,1} \leq \tilde{\lambda}_{R,S,2} \leq \ldots \leq \tilde{\lambda}_{R,S,r}\,, \IEEEyessubnumber \label{EqIneqConst1P3b}\\
& \text{(\ref{EqSrcPwrConst}) and (\ref{EqEqConstP3a})}\,.\nonumber
\end{IEEEeqnarray}
Problem P3(b) is convex and can be solved by an optimization solver. Analytical solutions are still attractive due to its low complexity. The challenge in deriving a closed-form solution is the ordering constraint in (\ref{EqIneqConst1P3b}). We find that when the ordering constraint in (\ref{EqIneqConst1P3b}) is relaxed, the output $\tilde{\lambda}_{R,S,m}^\star$ can still be in an increasing order if $\tilde{\lambda}_{R,S,m}$ in (\ref{EqSrcPwrConst}) are uniformly weighted (otherwise, the Lagrange multiplier for (\ref{EqSrcPwrConst}) would be non-uniformly weighted in the derived closed-form solution, which may violate the ordering of $\mathbf{\tilde{\lambda}}_{R,S}$). Therefore, we reformulate problem P3(b) as P3(c) by only replacing constraints (\ref{EqIneqConst1P3b}) and (\ref{EqSrcPwrConst}) with $\tilde{\lambda}_{R,S,1}, \tilde{\lambda}_{R,S,2}, \ldots, \tilde{\lambda}_{R,S,r} > 0$ and $\sum_{m = 1}^r \max\{\|\mathbf{h}_{e,m}\|^2\} \tilde{\lambda}_{R,S,m} \leq P_S$, respectively. The $\max\{\|\mathbf{h}_{e,m}\|^2\}$ is denoted as $h_{e,\text{max}}^2$ in subsequent parts. Problem P3(c) is convex, and KKT conditions are given by
\begin{IEEEeqnarray}{c}
\tilde{\lambda}^{\star}_{R,S,m} > 0\,,\quad m = 1,\ldots,r \label{EqKKT1P3c}\\
\sum_{m = 1}^r \tilde{\lambda}^{\star}_{R,S,m} \leq P_S/h_{e,\text{max}}^2\,,\label{EqKKT2P3c}\\
\sum_{m = 1}^r \left(\lambda_{f,m}(1 - \rho) - \rho\right)\tilde{\lambda}^{\star}_{R,S,m} = - \sum_{m = 1}^r (\sigma_n^2 + \beta_m)\lambda_{f,m} \nonumber\\
+ \rho P_D \lambda_{D,R,\text{max}}\,,\label{EqKKT3P3c}\\
\gamma^{\star}_{1,m} \geq 0\,,\label{EqKKT4P3c}\\
\gamma^{\star}_{1,m}\tilde{\lambda}^{\star}_{R,S,m} = 0\,,\label{EqKKT5P3c}\\
\gamma^{\star}_2 \geq 0\,,\label{EqKKT6P3c}\\
\gamma^{\star}_2 \left( \sum_{m = 1}^r \tilde{\lambda}^{\star}_{R,S,m} - \frac{P_S}{h_{e,\text{max}}^2} \right) = 0\,,\label{EqKKT7P3c}
\end{IEEEeqnarray}
and
\begin{equation}
\label{EqKKT8P3c}
- \frac{1}{\tilde{\lambda}^{\star}_{R,S,m}} - \gamma^{\star}_{1,m} + \gamma^{\star}_2 + \mu^{\star}\left(\lambda_{f,m}(1 - \rho) - \rho\right) = 0,
\end{equation}
where $\gamma^{\star}_{1,m}$, $\gamma^{\star}_2$, and $\mu^{\star}$ denote the optimal Lagrange multipliers. Eq. (\ref{EqKKT1P3c}), (\ref{EqKKT4P3c}), and (\ref{EqKKT5P3c}) reveal that $\gamma^{\star}_{1,m} = 0$. If $\gamma^{\star}_2 = 0$, according to (\ref{EqKKT8P3c}), it is obtained that
\begin{equation}
\label{EqClosedSolu1P3c}
\tilde{\lambda}^{\star}_{R,S,m} = \frac{1}{\mu^{\star} \left(\lambda_{f,m}(1 - \rho) - \rho\right)}\,,
\end{equation}
where $\mu^{\star}$ is obtained by solving $r / \mu^{\star} = \rho P_D \lambda_{D,R,\text{max}} - \sum_{m = 1}^r (\sigma_n^2 + \beta_m)\lambda_{f,m}$. Since $\tilde{\lambda}^{\star}_{R,S,m} > 0$ $\forall m$ and $\mu^{\star}$ also conforms to (\ref{EqKKT2P3c}), (\ref{EqClosedSolu1P3c}) is obtained provided
\begin{equation}
\label{EqP3cSolutionCond1}
\begin{cases}
\lambda_{f,m}(1 - \rho) - \rho \gtrless 0\,,\quad \forall m\\
0 \lessgtr \frac{1}{\mu^{\star}} \lesseqgtr \frac{P_S/h_{e,\text{max}}^2}{\sum^r_{m = 1}\frac{1}{\lambda_{f,m}(1 - \rho) - \rho}}
\end{cases}
\end{equation}
is satisfied. On the other hand, if $\lambda_3 > 0$, the optimal $\tilde{\lambda}^{\star}_{R,S,m}$ is achieved by
\begin{equation}
\label{EqClosedSolu2P3c}
\tilde{\lambda}^{\star}_{R,S,m} = \frac{1}{\gamma^{\star}_2 + \mu^{\star} \left(\lambda_{f,m}(1 - \rho) - \rho\right)}\,,
\end{equation}
where $\gamma^{\star}_2$ and $\mu^{\star}$ can be obtained by solving the non-linear system composed of (\ref{EqKKT2P3c}) and (\ref{EqKKT3P3c}).

\begin{algorithm}
\caption{Two-phase relaying with P3(b)}\label{AlgTwoPhaseRelayingP3b}
\begin{algorithmic}[1]
\State \textbf{Initialize} $\mathbf{\lambda}_f^{(0)}$ and $\mathbf{\tilde{\lambda}}_{R,S}^{(0)}$
\Repeat
    \State Update $\mathbf{\lambda}_f^{(\kappa + 1)}$ by calculating (\ref{EqClosedSoluP3a});
    \State Update $\mathbf{\tilde{\lambda}}_{R,S}^{(\kappa + 1)}$ by solving P3(b);
    \State $\kappa \gets \kappa + 1$;
\Until{\left|C(\mathbf{\lambda}_f^{(\kappa + 1)}, \mathbf{\tilde{\lambda}}_{R,S}^{(\kappa + 1)}) - C(\mathbf{\lambda}_f^{(\kappa)}, \mathbf{\tilde{\lambda}}_{R,S}^{(\kappa)})\right|<\epsilon}
\end{algorithmic}
\end{algorithm}
\begin{algorithm}
\caption{Two-phase relaying with P3(c)}\label{AlgTwoPhaseRelayingP3c}
\begin{algorithmic}[1]
\State \textbf{Initialize} $\mathbf{\lambda}_f^{(0)}$ and $\mathbf{\tilde{\lambda}}_{R,S}^{(0)}$
\Repeat
    \State Update $\mathbf{\lambda}_f^{(\kappa + 1)}$ by calculating (\ref{EqClosedSoluP3a});
    \If {\text{(\ref{EqP3cSolutionCond1}) is satisfied}}
        \State Update $\mathbf{\tilde{\lambda}}_{R,S}^{(\kappa + 1)}$ by calculating (\ref{EqClosedSolu1P3c});
    \Else
        \State Update $\mathbf{\tilde{\lambda}}_{R,S}^{(\kappa + 1)}$ by calculating (\ref{EqClosedSolu2P3c});
    \EndIf
    \State $\kappa \gets \kappa + 1$;
\Until{\left|C(\mathbf{\lambda}_f^{(\kappa + 1)}, \mathbf{\tilde{\lambda}}_{R,S}^{(\kappa + 1)}) - C(\mathbf{\lambda}_f^{(\kappa)}, \mathbf{\tilde{\lambda}}_{R,S}^{(\kappa)})\right|<\epsilon}
\end{algorithmic}
\end{algorithm}

As a summary, the proposed AO-based joint optimization algorithms are outlined in Algorithms \ref{AlgTwoPhaseRelayingP3b} and \ref{AlgTwoPhaseRelayingP3c}, where the objective function (\ref{C_NonIF_Scalar_P2}) is denoted as $C(\mathbf{\lambda}_f, \mathbf{\tilde{\lambda}}_{R,S})$. Since the separated optimization problems at the relay and the source (i.e. P3(b) or P3(c)) are convex problems with strictly convex objective functions, value of $C(\mathbf{\lambda}_f, \mathbf{\tilde{\lambda}}_{R,S})$ monotonically decreases with each iteration. Besides, the objective function (\ref{C_NonIF_Scalar_P2}) is lower-bounded. Thus, the two algorithms finally converge.

\section{Two-Phase Relaying without Energy Flow}
\label{SecTwoPhaseRelayingWoEF}
Considering that the relay only harvests power from the information flow (i.e., only the $S$-$R$ and the $R$-$D$ links exist in Fig. \ref{FigRelayingSchemes}), a two-phase relaying without energy flow is proposed. The design problem is formulated as
\begin{IEEEeqnarray}{cl}
\text{P4:}\, & \max_{\mathbf{Q}_S^{\prime}, \mathbf{F}^{\prime}} \frac{1}{2}\log\det \Big(\mathbf{I} + \nonumber\\
\IEEEeqnarraymulticol{2}{c}{(1\!-\!\rho)\mathbf{H}_{D, R} \mathbf{F}^{\prime} \mathbf{H}_{R, S} \mathbf{Q}_S^{\prime} \mathbf{H}_{R, S}^H \! \left[\mathbf{F}^{\prime}\right]^H \! \mathbf{H}_{D, R}^H \left[\mathbf{W}^{\prime}\right]^{-1} \Big)} \IEEEyessubnumber\\
\text{s.t.} & \text{Tr}\left\{ (1 - \rho) \mathbf{F}^{\prime} \mathbf{H}_{R, S} \mathbf{Q}_S^{\prime} \mathbf{H}_{R, S}^H \left[\mathbf{F}^{\prime}\right]^H + \sigma_n^2\mathbf{F}^{\prime}\left[\mathbf{F}^{\prime}\right]^H \right\}\nonumber\\
&= \nonumber \rho \text{Tr}\left\{\mathbf{H}_{R, S} \mathbf{Q}_S^{\prime} \mathbf{H}_{R, S}^H \right\}\,,\IEEEyessubnumber \label{FwdPowerConst_1P6}\\
&\text{Tr}\{\mathbf{Q}_S^{\prime}\} \leq P_S \,, \mathbf{Q}_S^{\prime} \succeq 0\,,\IEEEyessubnumber
\end{IEEEeqnarray}
where $\mathbf{W}^{\prime} = \sigma_{n}^2\mathbf{H}_{D, R} \mathbf{F}^{\prime} \left[\mathbf{F}^{\prime}\right]^H \mathbf{H}_{D, R}^H + \sigma_{n}^2\mathbf{I}$, and $\mathbf{F}^{\prime}$ denotes the relay processing matrix. Similar to the energy-flow-assisted two-phase relaying, channel diagonalization is also used to simplify the design problem. Different from the previous relaying scheme, due to the absence of the energy flow $\mathbf{Q}_D$, the $S$-$R$ channel can be decomposed by SVD. Recall that the SVD of $\mathbf{H}_{R, S} = \mathbf{U}_{R,S} \mathbf{\Sigma}_{R,S} \mathbf{V}_{R,S}^H$, where $\mathbf{\Sigma}_{R,S} = diag\{\lambda_{R,S,1}, \ldots, \lambda_{R,S,r}\}$; the EVD of $\mathbf{H}_{R, S} \mathbf{Q}_S^{\prime} \mathbf{H}_{R, S}^H = \mathbf{\tilde{U}}_{R,S}^{\prime} \mathbf{\tilde{\Sigma}}_{R,S}^{\prime} [\mathbf{\tilde{U}}_{R,S}^{\prime}]^H$, where $\mathbf{\tilde{\Sigma}}_{R,S}^{\prime} = diag\{\tilde{\lambda}_{R,S,1}^{\prime}, \ldots, \tilde{\lambda}_{R,S,r}^{\prime}\}$. Therefore, $\mathbf{Q}_S^{\prime} = \mathbf{V}_{R,S} \mathbf{\Sigma}_S^{\prime} \mathbf{V}_{R,S}^H$, where $\mathbf{\Sigma}_S^{\prime} = diag\{\lambda_{S,1}^{\prime}, \ldots, \lambda_{S,r}^{\prime}\}$, $\mathbf{\tilde{U}}_{R,S}^{\prime} = \mathbf{U}_{R,S}$, and $\mathbf{\tilde{\Sigma}}_{R,S}^{\prime} = \mathbf{\Sigma}_S^{\prime} \mathbf{\Sigma}_{R,S}$. For simplicity, we only focus on the case of uniform source power allocation, i.e. $\lambda_{S,m}^{\prime} = P_S/r$ $\forall m$. Assuming high receive SNR and omit the coefficient $1/2$, the power optimization of problem P4 is reformulated as
\begin{IEEEeqnarray}{ll}
\label{C_NonIF_Scalar_P7a}
\text{P5:}\, &\min_{\mathbf{\lambda}_f^{\prime}}  -\!\sum_{m = 1}^{r}\!\log\!\left(\frac{(1\!-\!\rho) \tilde{\lambda}_{R,S,m}^{\prime} \lambda_{f,m}^{\prime} \lambda_{D,R,m}}{\sigma_n^2\left(1\!+\!\lambda_{f,m}^{\prime} \lambda_{D,R,m}\right)}\right) \IEEEyessubnumber\\
\text{s.t.} & \lambda_{f,1}^{\prime}, \lambda_{f,2}^{\prime}, \ldots, \lambda_{f,r}^{\prime} \geq 0\,, \IEEEyessubnumber\\
& \sum_{m = 1}^r\!\left(\!(1\!-\!\rho)\!\lambda_{f\!,m}^{\prime} \tilde{\lambda}_{R,S,m}^{\prime}\!+\!\sigma_n^2\lambda_{f\!,m}^{\prime}\right) \!=\!\sum_{m = 1}^r\! \rho\tilde{\lambda}_{R\!,S\!,m}^{\prime}\IEEEyessubnumber \label{EqRelayPwrConstP7a}
\end{IEEEeqnarray}
where $\mathbf{\lambda}_f^{\prime} \triangleq [\lambda_{f,1}^{\prime}, \ldots, \lambda_{f,r}^{\prime}]^T$. The pairings of $\tilde{\lambda}_{R,S,m}^{\prime}$ and $\lambda_{D,R,m}$ for $m = 1,\ldots,r$ can be solved by Lemma \ref{Lemma1} with $P_D = 0$ and $\mathbf{\beta} = \mathbf{0}$. Hence, if $\tilde{\lambda}_{R,S,m}^{\prime} = P_S/r\cdot[\lambda_{R,S,\text{min}}, \ldots, \lambda_{R,S,\text{max}}]^T$ and $\lambda_{D,R,m}$ are arranged in an increasing order, the pairing problem is solved. Thus, the optimal $\lambda_{f,m}^{\prime}$ is obtained by
\begin{IEEEeqnarray}{l}
\label{EqClosedSoluP7a}
\left[\lambda_{f,m}^{\prime}\right]^{\star} = - \frac{1}{2\lambda_{D,R,m}}\nonumber\\
+ \frac{1}{2}\sqrt{\frac{1}{\lambda^2_{D,R,m}}\!+\!\frac{4}{\left[\nu^{\prime}\right]^{\star}\lambda_{D,R,m}\left((1 - \rho)\tilde{\lambda}_{R,S,m}^{\prime}\!+\! \sigma_n^2\right)}}\,,
\end{IEEEeqnarray}
where $\left[\nu^{\prime}\right]^{\star}$ is constrained by (\ref{EqRelayPwrConstP7a}).

\section{Simulation Results}
\label{SecPerformanceEvaluation}
\begin{figure}[!t]
\centering
\includegraphics[width = 2.2in]{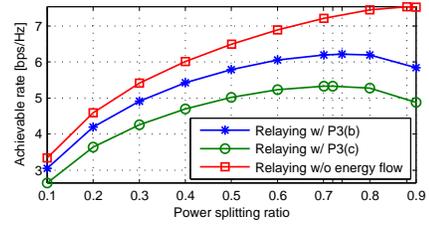}
\caption{Average rate as a function of PS ratio with $d_{DR}/d_{DS} = 0.65$.}
\label{Fig_Rate_as_a_func_of_rho}
\end{figure}
In the simulations, channel matrix $\mathbf{H}_{i,j}$ is generated by $\mathbf{H}_{i,j} = \Lambda_{i,j}^{-1} \mathbf{\bar{H}}_{i,j}$, where $\mathbf{\bar{H}}_{i,j}$ represents the small-scale fading. The large-scale fading is given by $\Lambda_{i,j}^{-1} = d_{ij}^{-3/2}$, where $d_{ij}$ is the distance between nodes $i$ and $j$ and $d_{DS} \triangleq d_{DR} + d_{RS}$. In the simulations, $r = 4$, $d_{DS} = 10$\,m, and  $\sigma_n^2 = 1$\,$\mu$W.

\begin{figure*}[t]
\centering
\subfigure[Average rate as a function of $d_{DR}/d_{DS}$ ratio with $P_D\!=\!0.5$\,W, $P_S\!=\!0.1$\,W.]{
\label{Fig_Rate_AR2ABratio_PA0p5_PB0p1_VarN1em6}
\includegraphics[width = 2.2in]
{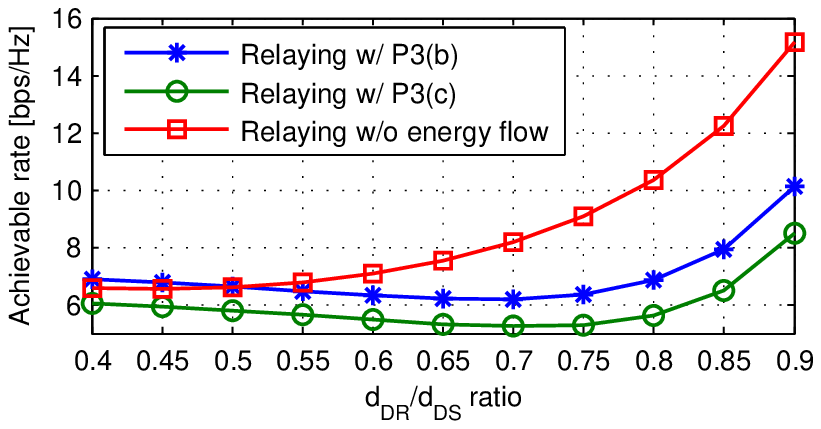}
}
\hfil
\subfigure[Best PS ratio as a function of $d_{DR}/d_{DS}$ with $P_D\!=\!0.5$\,W, $P_S\!=\!0.1$\,W.]{
\label{Fig_rho_AR2ABratio_PA0p5_PB0p1_VarN1em6}
\includegraphics[width = 2.2in]
{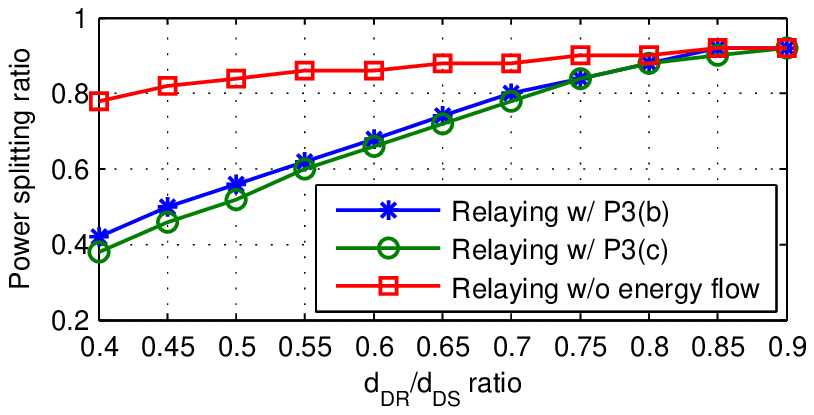}
}
\hfil
\subfigure[Average rate as a function of PS ratio with $P_D\!=\!5$\,W, $P_S\!=\!0.01$\,W.]{
\label{Fig_Rate_AR2ABratio_PA5_PB0p01_VarN1em6}
\includegraphics[width = 2.2in]
{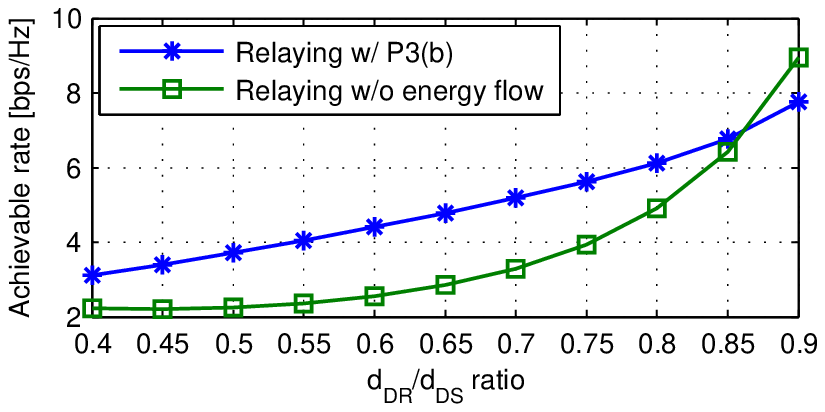}
}
\caption{Rate performance under different $d_{DR}/d_{DS}$ ratios. The PS ratio is exhaustively searched among 0.02:0.02:0.98 to maximize the average rate.}
\end{figure*}

Fig. \ref{Fig_Rate_as_a_func_of_rho} shows that for a certain $d_{DR}/d_{DS}$ value, the average achievable rate of the proposed two-phase relaying schemes firstly increases to a stationary point at PS ratios of 0.72, 0.74, and 0.88, respectively. Then, the rate decreases as the PS ratio increases. This is because when the PS ratio is small, less power is harvested for forwarding, which limits the receive SNR at $D$; when the PS ratio is large, less signal power remains for the ID receiver at $R$ and the SNR at $R$ decreases. Both the above two cases degrade the achievable rate. It is also observed that the two-phase relaying with P3(c) is always inferior to P3(b). The reason lies in that the modified source power constraint $\sum_{m = 1}^r \max\{|\mathbf{h}_{e,m}|^2\} \tilde{\lambda}_{R,S,m} \leq P_S$ in P3(c) would limit the received information signal power at the relay, and $S$-$R$ link performance is sacrificed.

Fig. \ref{Fig_Rate_AR2ABratio_PA0p5_PB0p1_VarN1em6} shows the average rate as a function of $d_{DR}/d_{DS}$ ratio, where an exhaustive search is performed to find the best PS ratio for each $d_{DR}/d_{DS}$ ratio. It is shown that the rate of the scheme without energy flow decreases as the relay approaches to $D$, because $R$ only extracts forwarding power from the information flow. Nevertheless, harvesting the energy from $D$, the rate of the energy-flow-assisted scheme can increase as $d_{DR}/d_{DS}$ decreases. However, it is observed that when $R$ is close to $S$ (e.g. $d_{DR}/d_{DS}\!=\!0.9$ where the energy flow slightly contributes to the forwarding power), the scheme without energy flow can outperform the energy-flow-assisted scheme. This is because to enhance the power usage effectiveness, the LSV of $\mathbf{\tilde{H}}_{R,S}$ is forced to be $\mathbf{V}_{D,R}^{\ast}$ with the HPM-PLM strategy, which makes the source beamforming matrix not always unitary. Fig. \ref{Fig_rho_AR2ABratio_PA0p5_PB0p1_VarN1em6} demonstrates that if $R$ is close to $D$, the $S$-$R$ link becomes the critical link; thus, a lower PS ratio is needed. Otherwise, the $R$-$D$ link becomes the critical link, and a higher PS ratio is needed. Compared with the energy-flow-assisted scheme, the scheme without energy flow needs higher PS ratios, because its relay power only comes from the information flow.

Fig. \ref{Fig_Rate_AR2ABratio_PA5_PB0p01_VarN1em6} studies the asymmetric scenario where the power budget at $D$ is increased while that at $S$ is decreased. It is observed that the rate of the energy-flow-assisted scheme increases as $R$ moves towards $S$. This is because with adequately large power budget at $D$, the receive SNR at $D$ can still increase, although the harvested power of the energy flow at $R$ decreases. Compared with Fig. \ref{Fig_Rate_AR2ABratio_PA0p5_PB0p1_VarN1em6}, the energy-flow-assisted scheme outperforms (rate-wise) the scheme without energy flow at most $d_{DR}/d_{DS}$ ratios due to the efficient utilization of the harvested power. Recall that the energy-flow-assisted scheme prefers a high-quality $S$-$R$ channel, e.g. a high transmit SNR at $S$. Although the transmit SNR at $S$ decreases in the case of Fig. \ref{Fig_Rate_AR2ABratio_PA5_PB0p01_VarN1em6}, the relative difference in rate at $d_{DR}/d_{DS}\!=\!0.9$ between the two schemes in Fig. \ref{Fig_Rate_AR2ABratio_PA5_PB0p01_VarN1em6} is smaller than that in Fig. \ref{Fig_Rate_AR2ABratio_PA0p5_PB0p1_VarN1em6}. This illustrates that the rate can benefit from the energy-flow-assisted scheme, when the power budget at $D$ is adequately larger than that at $S$.

\section{Conclusion}
\label{SecConclusion}
In this paper, we have investigated JWIPT relaying schemes in an AF MIMO one-way relay network, where a wireless-powered autonomous relay is deployed. Considering possible simultaneous transmission of energy and information, we have proposed an energy-flow-assisted two-phase relaying and a two-phase relaying without energy flow. Simulation results reveal that the rate can benefit from the energy-flow-assisted scheme if the transmitted power of the energy flow is adequately larger than that of the information flow. Otherwise, the scheme without energy flow would be better.

\bibliographystyle{IEEEtran}
\bibliography{IEEEabrv,BibPro}
\end{document}